\documentclass[%
preprint,
groupedaddress,
 amsmath,amssymb,
pra,
]{revtex4-1}

\usepackage{dcolumn}
\usepackage{bm}
\usepackage{color}
\usepackage{graphicx}
\usepackage{epstopdf}
\usepackage{rotating}
\usepackage{mathtools}
\usepackage{mathrsfs}  

\newcommand{\bra}[1]{\langle #1|}
\newcommand{\ket}[1]{|#1\rangle}
\newcommand{\braket}[2]{\langle #1|#2\rangle}
\def\Ham{{\cal H}}
\let\oldhat\hat
\renewcommand{\hat}[1]{\oldhat{\mathbf{#1}}}

\renewcommand{\vec}[1]{\mathbf{#1}}

\def\k{\vec{k}}

\newcommand{\be}{\begin{equation}}
\newcommand{\ee}{\end{equation}}

\begin{document}



\title{Strong ultrafast demagnetization due to the intraband
  transitions
}

\author{Mitsuko Murakami}\email[]{mitsuko@phys.ntu.edu.tw}
\affiliation{Department of Physics, Indiana State University, Terre Haute, IN 47809, USA}

\author{G.~P.~Zhang}\email[]{guo-ping.zhang@outlook.com. https://orcid.org/0000-0002-1792-2701}
\affiliation{Department of Physics, Indiana State University, Terre Haute, IN 47809, USA}

\date{\today}

\begin{abstract}
Demagnetization in ferromagnetic transition metals driven by a
femtosecond laser pulse is a fundamental problem in solid state
physics, and its understanding is essential to the development of
spintronics devices.  Ab initio calculation of time-dependent magnetic
moment in the velocity gauge so far has not been successful in
reproducing the large amount of demagnetization observed in
experiments.  In this work, we propose a method to incorporate
intraband transitions within the velocity gauge through a convective
derivative in the crystal momentum space.  Our results for
transition-element bulk crystals (bcc Fe, hcp Co and fcc Ni) based on
the time-dependent quantum Liouville equation show a dramatic
enhancement in the amount of demagnetization after the inclusion of an
intraband term, in agreement with experiments.  We also find that the
effect of intraband transitions to each ferromagnetic material is
distinctly different because of their band structure and spin property
differences.  Our finding has a far-reaching impact on understanding
of ultrafast demagnetization.
\end{abstract}


\maketitle


\section{\label{sec:intro} Introduction}
Ever since the pioneering experiment by Beaurepaire {\it et al.} in
1996, ultrafast demagnetization driven by a femtosecond laser pulse
has been one of the most-discussed problems in solid-state physics
\cite{ourbook}.  When irradiated with a 620-nm mode-locked dye laser
pulse of duration 60 fs, a 22-nm nickel film was found to reduce its
magnetization by more than 40\% in 260 fs \cite{eric}.  Before their
experiment, it was believed that the spin relaxation time in itinerant
ferromagnets is of the order of 100 ps
\cite{vaterlaus1991,koopmans2010}.  Subsequent experiments have
confirmed that the laser-induced demagnetization of ferromagnetic
transition metals (Co, Ni, Fe) indeed takes place within 500 fs (see
Refs.~\cite{rasingreview,mplb16,prl20} and references therein).
Implication of this femtomagnetism is profound and has lead to the
development of spintronic devices such as magnetic random access
memory \cite{kimel2019}.  Building a consensus among theorists has not
been easy, however; possible explanations include, e.g., Stoner
excitation \cite{scholl1997}, spin-orbit coupling \cite{prl00},
domain-wall motion \cite{gerrits2002,ramsay,lan2017}, transient
exchange-splitting reduction \cite{rhie2003,muller2009,prl00}, Faraday
rotation \cite{kimel2005}, Elliot-Yafet scattering
\cite{steiauf2009,krauss2009,koopmans2010}, magnon emission
\cite{schmidt2010,haag2014}, superdiffusive spin transport
\cite{battiato2010,schellekens2013}, ultrafast heating
\cite{ostler2012}, and Einstein-de Haas effect \cite{dornes2019}.
Some of these phenomena have similar time scales, which makes it
challenging to distinguish the prominent mechanism
\cite{turgut2016,choi2017,buhlmann2018,scheid2021}.

In the last several years, ab initio calculations of the ultrafast demagnetization of ferromagnetic materials were made, based on the time-dependent density functional theory (TDDFT) \cite{krieger2015,elliott2016b,siegrist2019,scheid2021,dewhurst2021b} or the time-dependent Liouville density functional theory \cite{jpcm16}.  There was a common problem in both approaches, that is, the amount of demagnetization turned out too small to be compared with experiments.  For a typical fluence of femtosecond lasers ($\sim$10 mJ/cm$^2$), these ab initio calculations predicted less than 10\% reduction in the magnetic moments, whereas complete demagnetization of a pure-crystal Ni film has been reported experimentally at a moderate fluence of 4.43 mJ/cm$^2$ \cite{you2018}.    One likely explanation to this discrepancy between ab initio calculations and experiments is the effect of intraband transitions in the crystal momentum space \cite{aversa1995,golde2008}.  Time evolution of a state vector driven by a laser field can be seen as a sequence of dipole transitions between the laser-accelerated Bloch states.   Intraband transitions connect those Bloch states which are in the same band but of different crystal momenta $\k$, whereas interband transitions connect those in different bands at the same $\k$. Their transition probabilities are proportional to dipole matrix elements between the Bloch states, but ease of calculation depends on the choice of gauge.  Although intra- and interband transition dipole moments are readily distinguishable in the length gauge, calculation of position vector $\langle\vec{r}\rangle$ for intraband transitions in particular involves $-i\nabla_\k\delta(\k-\k')$, which is highly singular, as well as Berry connections ${\cal A}=\langle-i\nabla_\k\rangle$ which are not uniquely determined in general \cite{blount1962}.  Calculation of the dipole moment from a momentum vector $\langle \vec{p}\rangle$ in the velocity gauge, on the other hand, is more straightforward but does not allow a separation of intraband and interband components.  An alternative approach that distinguishes the inter- and the intraband dipole moments is to use time-dependent basis functions (Houston states) rather than Bloch states in the time evolution \cite{krieger1986}, but unlike the velocity-gauge method, the Houston state treatment is numerically unstable when more than a few low-lying energy bands are used \cite{wu2015}. 

In 2009, our velocity-gauge calculation based on the time-dependent {quantum Liouville equation (QLE)} in Ref.~\cite{np09} was able to show that the time evolutions of magnetic and dipole moments are correlated, settling a decade-long debate as to whether measurements based on the time-resolved magneto-optical Kerr effect (TRMOKE) could really probe magnetization \cite{koopmans2000}.  Yet, which type of transitions, intra- or interband, is more significant to femtomagnetism remains unknown.  Meanwhile, high harmonic generation from bulk crystals (ZnO, GaSe, SiO$_2$)  in the terahertz, midinfrared, and ultimately the extreme ultraviolet (EUV) region  has been reported since 2011 \cite{ghimire2011,schubert2014,luu2015}.  Calculations of solid-state high harmonic spectra followed, which were obtained from the square modulus of Fourier-transformed expectation values of the dipole moment, and showed that the contribution from interband transitions is several orders of magnitude more intense than from intraband transitions for those harmonics whose energy is beyond the minimum band-gap \cite{vampa2014,wu2015,vampa2015}, although EUV high harmonics of SiO$_2$ were claimed to be from the intraband dipole only \cite{luu2015}.  Similar investigations are needed for the complete understanding of femtomagnetism.

\begin{figure*}[!h]
\centering
\includegraphics[width=.9\textwidth, trim = 0in 0in 0in 0in, clip]{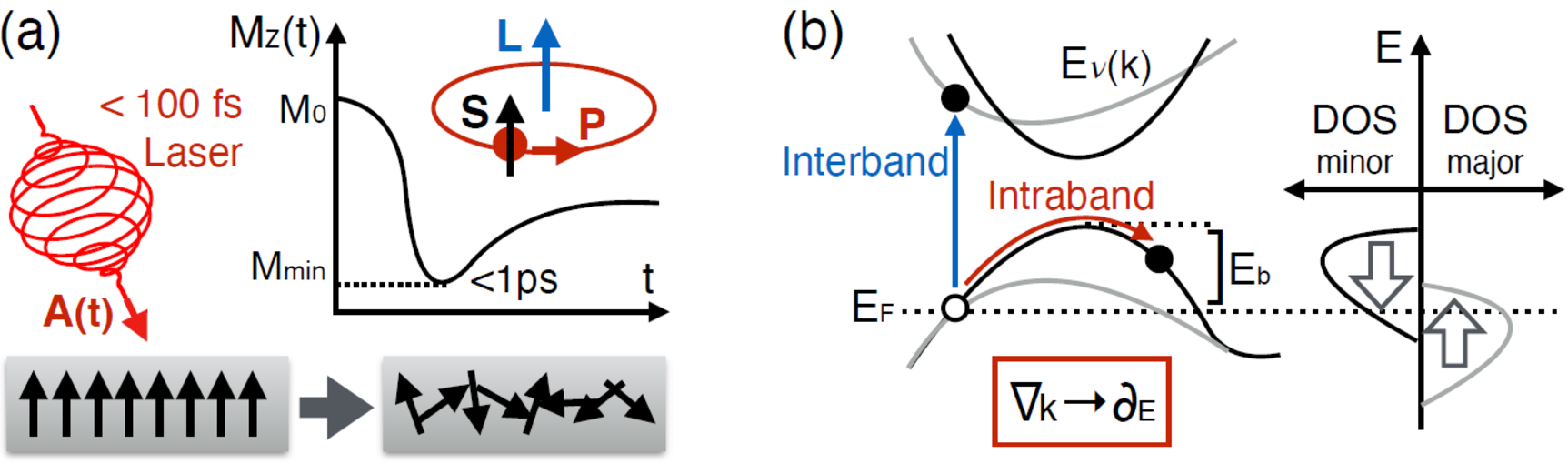}
\caption{(Color online)  (a) Ultrafast demagnetization of ferromagnets using a femtosecond laser pulse.  Demagnetization takes place in less than a picosecond, followed by slower remagnetization during which electron temperature cools off via electron-phonon scattering.  (b)  Electron dynamics in the crystal momentum representation can be described as a sequence of dipole transitions between the accelerated Bloch states of crystal momentum $\k(t)=\k(0)-\vec{A}(t)$.  The driving laser causes both inter- and intraband transitions above the Fermi energy $E_{\rm F}$, which equilibrate the density of states (DOS) between major and minority spins.  Calculation of intraband transitions involves the gradient of electron wave functions with respect to crystal momentum vector $\k$, which can be replaced with derivatives with respect to energy.  To assess the amount of intraband dipole transitions contributing to demagnetization, we can restrict the change in energy within a bracket energy $E_b$ and see how it affects the loss of magnetization. }
\label{fig:scheme}
\end{figure*}

In this paper, we study the ultrafast demagnetization of ferromagnetic
solids of transition elements (bcc Fe, hcp Co and fcc Ni) by solving
the {time-dependent QLE}.  {The velocity gauge is used, but an effect
of intraband transition is added using the convective derivative
\cite{kittel2,mahan2000,griffiths2013}.}  The key idea is to replace the
troublesome $\k$-gradient operator in the convective derivative with
the energy derivative, which significantly simplifies the treatment of
intraband transitions.  In order to assess the effect of intraband
transitions on demagnetization, we introduce a parameter called the
bracket energy in our calculation, which restricts the amount of
energy an electron could absorb from intraband transitions
(Fig.~\ref{fig:scheme}).  Our results show that including an intraband
transition term in the velocity-gauge {QLE} leads to a dramatic
enhancement in the amount of demagnetization in all three
ferromagnets, in accordance with experiments.

The paper is organized as follows.  
In  Sec.~\ref{sec:methods}, we present our method and the rationale behind the implementation of intraband transitions in the velocity gauge.
Results and discussions are presented in Sec.~\ref{sec:results}, followed by our conclusion in Sec.~\ref{sec:conc}.  Atomic units ($m=e=\hbar=1$) are used throughout, unless otherwise noted.

\section{\label{sec:methods}Methods}

\subsection{Density Functional Theory}
Let $\ket{\nu,\k}$ be the Bloch state of energy $E_{\nu,\k}$, where $\nu$ and $\k$ are the band index and the crystal momentum, respectively.  It is the solution of the stationary Kohn-Sham equation 
\begin{equation}\label{eq:KS}
\Ham_0 \ket{\nu,\k} = E_{\nu,\k}  \ket{\nu,\k},
\end{equation}
where $\Ham_0$ includes the single-electron kinetic energy operator and Kohn-Sham potential \cite{koelling1977,perdew2003}.  The Kohn-Sham potential is a functional of the electron density $n(\vec{r})$, given by 
\begin{equation}\label{eq:n1}
n(\vec{r}) = \sum_{\nu,\k} \Theta(E_{\rm F}-E_{\nu,\k})|\psi_{\nu,\k}(\vec{r})|^2,
\end{equation}
where $\psi_{\nu,\k}(\vec{r})=\braket{\vec{r}}{\nu,\k}$ is the coordinate representation of the Bloch states, and the Fermi energy $E_{\rm F}$ is chosen to satisfy
\begin{equation}
\dfrac1\Omega \int_\Omega n(\vec{r}) d^3 r = N,
\end{equation}
in which $N$ is the number of electrons in a unit cell of volume $\Omega$.  The step function $\Theta(E_{\rm F}-E_{\nu,\k})$ in Eq.~(\ref{eq:n1}) ensures that all Kohn-Sham spin orbitals with energy $E_{\nu,\k} \leq E_{\rm F}$ are singly occupied, and empty otherwise. 

We solve a set of Kohn-Sham equations (\ref{eq:KS}) iteratively by using the full-potential linearized augmented plane-wave (FLAPW) method implemented in the WIEN2K software \cite{blaha2010,blaha2020}.  As for the exchange-correlation term in the Kohn-Sham potential, the Perdew-Burke-Ernzerhof (PBE) functional based on the generalized gradient approximation (GGA) is used \cite{pbe}.  The spin-orbit coupling is included using a second-variational method in the same self-consistent iteration \cite{macdonald1980}.  All of our eigenstates are spin-mixed and have two components.

\subsection{Quantum Liouville Equation}

The time-dependent Kohn-Sham orbital for the $i$th electron ($i=1,2,\cdots N$) is expanded in the basis of Bloch states as
\begin{equation}
\ket{\psi_i(t)} = \sum_{\nu,\k} c^i_{\nu,\k}(t) \ket{\nu,\k},
\end{equation}
with the normalization condition: $\braket{\psi_i(t)}{\psi_i(t)}=1$, which in light of the orthonormality $\braket{\nu\k}{\nu'\k'}=\delta_{\nu\nu'}\delta_{\k,\k'}$ of Bloch states implies
\begin{equation}
\sum_{\nu,\k} |c^i_{\nu,\k}(t)|^2 = 1.
\end{equation}
The expansion coefficients $c^i_{\nu,\k}(t)=\braket{\nu,\k}{\psi_i(t)}$ are then called the crystal momentum representation of a state vector \cite{blount1962}.  

The density operator of Kohn-Sham orbitals is defined by 
\begin{equation}
\rho(t)\equiv\sum_i \ket{\psi_i(t)}\bra{\psi_i(t)},
\end{equation}
whose matrix elements in the Bloch basis are 
\begin{equation}
\rho_{\nu\nu'}^\k(t)=\bra{\nu\k}\rho(t) \ket{\nu'\k}= \sum_i c^{i \;\dagger}_{\nu,\k}(t) c^i_{\nu',\k'}(t).
\end{equation}
The initial state of the density matrix is given by
\begin{equation}\label{eq:init}
\rho_{\nu\nu'}^\k(0) = \delta_{\nu\nu'}\Theta(E_{\rm F} - E_{\nu,\k}),
\end{equation}
which is consistent with the ground-state electron density given by Eq.~(\ref{eq:n1}).  

We evolve the density operator (\ref{eq:init}) by solving the time-dependent quantum Liouville equation {(QLE)} in the velocity gauge; that is, 
\begin{equation}\label{eq:QLE}
i\hbar\dfrac{\partial \rho}{\partial t} = [\Ham_0+\Ham_I^{\rm VG},\rho(t)].
\end{equation}
 In the limit of dipole approximation, the interaction Hamiltonian in the velocity gauge is \cite{madsen2002,baurer2005}
\begin{equation}\label{eq:VG}
\Ham_I^{\rm VG} = - \vec{p}\cdot\vec{A}(t),
\end{equation}
where $\vec{A}(t)$ is the vector potential of a circularly-polarized driving laser field of a Gaussian envelope on the $xy$-plane, given by
\begin{equation}
\vec{A}(t) = A_0 e^{-t^2/\tau^2} \left[ 
(\cos\omega t) \hat x \pm (\sin \omega t )\hat y\right],
\end{equation}
where $\tau$ is the laser pulse duration.  
A linearly polarized driving laser pulse can similarly be implemented as well.  Accordingly, Eq.~(\ref{eq:QLE})  in the matrix-element form becomes  \cite{prb09}

\begin{equation}\label{eq:QLE2}
i\hbar\dfrac{\partial \rho_{\nu\nu'}^\k}{\partial t} = (E_{\nu,\k}- E_{\nu',\k}) \rho_{\nu\nu'}^\k
-\vec{A}(t)\cdot \sum_\mu \left( \vec{p}_{\nu\mu}^\k \rho_{\mu\nu'}^\k - \rho_{\nu\mu}^\k \vec{p}_{\mu\nu'}^\k \right),
\end{equation}
where $\k$ is 
 time-independent. 
$\vec{p}^\k_{\nu\nu'}=\bra{\nu,\k}-i\nabla_\vec{r}\ket{\nu',\k}$ is
the momentum matrix element between  energy bands $\nu$
and $\nu'$ is calculated within a module of the WIEN2K code using the LAPW basis \cite{draxl2006}.  
 Then, the expectation value of a magnetization is evaluated at each timestep as 
\begin{equation}
\langle \vec{M}(t) \rangle = \sum_{\k} {\rm Tr}(\rho^\k(t) \bm{S}^\k),
\end{equation}
where $\bm{S}^\k=S_x^\k\hat{x}+S_y^\k\hat{y}+S_z^\k\hat{z}$ is the spin matrix vector in the LAPW basis \cite{prb09}.

There is an important advantage of the velocity gauge over the length gauge in the crystal momentum representation that, as long as the dipole approximation holds and electron scattering is neglected,  $\Ham_I$ induces no coupling between different values of crystal momentum $\k$.  This is because of the Bloch theorem:
\begin{equation}
\psi_{\nu,\k}(\vec{r}+\vec{R}_m) = e^{i\k\cdot\vec{R}_m}\psi_{\nu,\k}(\vec{r}),
\end{equation}
where $\vec{R}_m$ are the translational lattice vectors
($m=1,2,\cdots$), and of the fact that the momentum operator
$\vec{p}=-i\nabla_\vec{r}$ is invariant of space translation.  As a
result, the transition moment in the velocity gauge couples two Bloch
states as \cite{korbman2013}
\begin{equation}
\bra{\nu,\k} \vec{p} \ket{\nu',\k'} 
= \sum_m  e^{-i(\k-\k')\cdot\vec{R}_m} 
 \int_\Omega  \psi^*_{\nu,\k}(\vec{r}) \left(-i\nabla_{\vec r}
 \psi_{\nu',\k'}(\vec{r}) \right) d {\bf r} 
\; \propto  \delta_{\k,\k'},
\end{equation}
in which the integral is taken over a unit cell with $\vec{R}_m=0$.
On the other hand, the position operator $\vec{r}$ in the length gauge
does not conserve the crystal momentum $\k$, and the transition dipole
matrix elements involve the off-diagonal
Berry connections ${\cal A}_{\nu,\nu'}^{\k,\k'} \equiv \bra{\nu,\k} -i\nabla_\k \ket{\nu',\k'}$  \cite{aversa1995,cheng2019}.

It is insightful to note that the {QLE} in the crystal momentum representation (\ref{eq:QLE2}) becomes the semiconductor Bloch equations \cite{golde2008,luu2016} if the density matrix elements are given by $\rho_{\nu\nu'}^\k\equiv \langle a_{\nu,\k}^\dagger a_{\nu',\k}\rangle$, where $a_{\nu,\k}^\dagger$ and $a_{\nu,\k}$ are creation and annihilation operators for the Bloch states, diagonal terms ($\nu=\nu'$) describing the population and off-diagonal terms ($\nu\ne\nu'$) the coherence \cite{al-naib2014}.

\subsection{Treatment of intraband transition}

The QLE (Eq. \ref{eq:QLE2}) in the velocity gauge using the
time-independent $\k$ in the previous section has a
deficiency. Consider the system only has a single band, such as the
nearly free-electron model.  Equation (\ref{eq:QLE2}) is reduced to
\begin{equation}
i\hbar\dfrac{\partial \rho_{\nu\nu}^\k}{\partial t} = (E_{\nu,\k}- E_{\nu,\k}) \rho_{\nu\nu}^\k
-\vec{A}(t)\cdot  \left( \vec{p}_{\nu\nu}^\k \rho_{\nu\nu}^\k -
\rho_{\nu\nu}^\k \vec{p}_{\nu\nu}^\k \right) =0.
\label{1band}
\end{equation}
This demonstrates that if we ignore the time dependence in ${\bf k}$,
the density in the velocity gauge remains unchanged, so is any
observable, which is clearly invalid.

 Physically, even a static field $\vec{F}$ moves one electron from
 $\vec{k}_1$ to $\vec{k}_2$.  To see this, suppose that our
 one-dimensional system is along the $x$ axis. The crystal momentum
 $k_x(t)$ at time $t$ is connected with the initial ${k}_x(0)$
 through \cite{kittel}
\begin{equation}
{k}_x(t) = {k}_x(0)+ {F}t/\hbar.
\end{equation}
In the velocity gauge, the spatial change is now translated to the
crystal momentum $\vec{k}$ change in time as
$\vec{k}(t)=\vec{k}(0)-\vec{A}(t)$.

The situation resembles the
classical Boltzmann equation, which, besides the time-dependence,
 depends on both the position and velocity \cite{kittel}:
\begin{equation}
\dfrac{\partial f}{\partial t} + \vec{v}\cdot\nabla_{\vec{r}}f + \dfrac{d\vec{v}}{dt}\cdot\nabla_{\vec{v}}f = \left( \dfrac{\partial f}{\partial t} \right)_{\rm coll},
\end{equation}
where  $f$ is the distribution function and 
the right hand side is due to collisions. Since the diagonal
element of a density matrix is just the distribution used in the
Boltzmann equation, one can see the direct connection between the
quantum and classical mechanics and the result must agree with each
other a single band limit.

Now we  can understand how Eq. \ref{1band}
fails. Although its time-derivative is zero, $\rho$ is changed through
$\rho^{\k(t)}$. A simple solution might be to change every $\k$ to
$\k(t)$ in Eq. \ref{eq:QLE2}, but it would introduce two
difficulties. First, Eq.  \ref{eq:QLE2} becomes a time-dependent {\it
  functional} differential equation, unsolvable in general. Second,
the pure intraband transition has
no memory effect.  Once the laser field is gone, $\rho$
returns to its original value, so the intraband transition alone cannot
describe  demagnetization and we
must include interband transitions.
A compromise is to rewrite  $d\rho^{\k(t)}/dt = \partial \rho^{\k(0)}/dt + \partial
\rho/\partial \k \cdot \partial \k/\partial t$, where  $\rho^{\k(0)}$
is the density for the time-independent $\k$. Therefore,
the partial
time-derivative on the left hand side of Eq.~(\ref{eq:QLE2}) becomes
the convective derivative as \cite{kittel2,mahan2000,griffiths2013}
\begin{equation}\label{eq:conv}
{\dfrac{d}{dt}} \longrightarrow \dfrac{\partial}{\partial t}  + \vec{v}_k \cdot\nabla_\k .
\end{equation}
where $\vec{v}_k \equiv \partial \k/\partial t$.  Because $\k(t) = \k(0)-\vec{A}(t)$, we write \cite{krieger1986}
\begin{equation}
\vec{v}_k = - \dfrac{\partial \vec{A}}{\partial t} = \vec{E}(t).
\end{equation}

The derivative with respect to the crystal momentum in the convective derivative is numerically challenging, and it is often obtained through Wannier functions \cite{blount1962}.  We shall instead rewrite the $\k$-gradient of Eq.~(\ref{eq:conv}) in terms of band energy $E_{\nu,\k}$ as
\begin{equation}\label{eq:kderiv}
\nabla_\k = \dfrac{\partial}{\partial \k} 
= \dfrac{ \partial E_{\nu,\k}}{\partial \k} \dfrac{\partial}{\partial E_{\nu,\k}} = \vec{p}^\k_{\nu\nu} \dfrac{\partial}{\partial E_{\nu,\k}},
\end{equation}
where $\vec{p}^\k_{\nu\nu}=\partial E_{\nu,\k}/\partial \k$ are the group velocity of an electron in the $\nu$th band \cite{ashcroft}.  Then, we can write
\begin{equation}
\vec{v}_\k\cdot\nabla_\k= \vec{E}(t) \cdot \vec{p}^\k_{\nu\nu} \dfrac{\partial}{\partial E_{\nu,\k}}.
\end{equation}
Therefore, the equation (\ref{eq:QLE2}) becomes
\begin{equation}\label{eq:QLE3}
i\hbar\dfrac{\partial \rho_{\nu\nu'}^\k}{\partial t} = (E_{\nu,\k}- E_{\nu',\k}) \rho_{\nu\nu'}^\k
-\vec{A}(t)\cdot \sum_\mu \left( \vec{p}_{\nu\mu}^\k \rho_{\mu\nu'}^\k - \rho_{\nu\mu}^\k \vec{p}_{\mu\nu'}^\k \right)
-\vec{E}(t) \cdot \vec{p}^\k_{\nu\nu} \delta_{\nu\nu'} \dfrac{\partial\rho_{\nu\nu'}^\k} {\partial E_{\nu,\k}}.
\end{equation}
 Therefore, the energy derivative of diagonal elements (which account for the intraband transition) in the last term of Eq.~(\ref{eq:QLE3}) is calculated as
\begin{equation}\label{eq:Ederiv}
\dfrac{\partial\rho_{\nu\nu}^\k} {\partial E_{\nu,\k}} 
\approx
\sum_\mu \Theta(E_b-|E_{\mu,\k}-E_{\nu,\k}|) \frac{\rho_{\mu,\mu}^\k(t)-\rho_{\nu,\nu}^\k(t)} {|E_{\mu,\k}-E_{\nu,\k}|},
\end{equation}
where the step function assures that the summation over $\mu$ is zero
unless the energy difference between $E_{\mu,\k}$ and $E_{\nu,\k}$ is
within a bracket energy $E_b$.  The bracket energy is a parameter used
in our calculation to control the amount of intraband transitions. 

We should note that expressing the $\k$-gradient operator as a product of the group velocity and the energy derivative as in Eq.~(\ref{eq:kderiv}) is not a new idea, for it has similarly been used in the transport theory which assumes the electron number density to be \cite{peierls}
\begin{equation}
n(\k,E) = f(E) + n^{[1]}(\k),
\end{equation}
that is, the $\k$-dependence is simply the first-order correction to the Fermi distribution $f(E)$.  
Under the steady state ($\partial n/ \partial t = 0$) and the closed circuit ($\nabla n = 0$) conditions, the above assumption leads to 
\begin{equation}
\dfrac{n^{[1]}(\k)-\bar{n}^{[1]}}{\tau} = -\vec{E}(t)\cdot\vec{p} \dfrac{\partial f}{\partial E},
\end{equation}
where $\tau$ is so-called the transport relaxation time, and $\bar{n}^{[1]}$ is a constant. 
  This
intraband transition provides a basis for 
 a possible demagnetization mechanism in the crystal momentum space, just as the superdiffusive spin transport in real space has been proposed phenomenologically \cite{battiato2010}.

Our present calculation does not include the effect of electron-phonon scattering, which could be incorporated by adding a phenomenological dephasing term in Eq.~(\ref{eq:QLE3}) \cite{luu2016,cheng2019}.   In general, the electron-phonon scattering leads to remagnetization after approximately 1 ps, due to the equilibration of electron temperature with phonons \cite{roth2012}.  Since we compare our results with experiments at a shorter time window before remagnetization starts around 1 ps, however, adding the dephasing term is not necessary.   
Equation (\ref{eq:QLE3}) provides a way to include an intraband term
in the {QLE} using the velocity gauge, which is simpler than a prior
study \cite{cheng2019}.

\section{\label{sec:results}Results and Discussions}

Our prior study based on TDDFT demonstrated that the electron correlation and the memory kernel frequency dependence in a density functional could bring a strong demagnetization \cite{prl00}.  
The net effect is to build in more spin flipping contribution.  This motivated us to wonder whether the intraband transition can do the same job.  However, our prior study did not include the intraband transition.

\subsection{Effects of intraband transitions}

\begin{figure}[!h]
\centering
\includegraphics[width=.5\textwidth, trim = 0in 0in 0in 0in, clip]{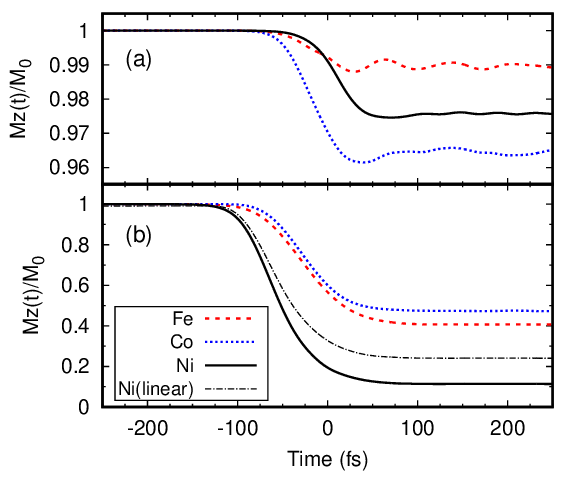}
\caption{(Color online)  Normalized magnetization of bcc Fe, hcp Co and fcc Ni as a function of time, when the bracket energy $E_b$ in Eq.~(\ref{eq:Ederiv}) is set to (a) zero, i.e., there is no intraband transition, and (b) their cutoff values (1 eV, 3 eV and 1.5 eV, respectively) to achieve the theoretically maximum magnetization reduction.  The 1.6 eV (775 nm) driving laser pulse is circularly polarized and has a vector potential of duration $\tau=60$ fs and a peak strength of  $A_0=0.05$ V$\cdot$fs/{\AA}, whose corresponding fluence is $F_0= \sqrt{\frac{\pi}{2}}\frac{c\epsilon_0}{2}(\omega A_0)^2 =14.7$ mJ/cm$^2$. Also shown in (b) with a thin dash-dotted line is a result for fcc Ni using a linearly-polarized laser with the same duration and peak strength.  We see that the demagnetization by a linearly-polarized laser is not as effective as a circularly-polarized one.}
\label{fig:spin}
\end{figure}

In this section, we study the demagnetization of three ferromagnetic materials, bcc Fe, hcp Co and fcc  Ni, driven by a circularly-polarized, $\omega=1.6$ eV (775 nm) laser pulse of duration  $\tau=60$ fs.  
In Fig.~\ref{fig:spin}, we plot the time profile of normalized magnetization of three ferromagnetic crystals, bcc Fe, hcp Co and fcc Ni, driven by a circularly-polarized, 1.6 eV, 60 fs laser pulse of peak vector potential $A_0=0.05$ V$\cdot$fs/{\AA}, when the bracket energies $E_b$ in Eq.~(\ref{eq:Ederiv}) are set to (a) zero, i.e., there are no intraband
transitions, or (b) the cutoff values ($E_b=1$ eV, 3 eV and 1.5 eV, respectively) which achieve the maximum demagnetization.   Different $E_b$ is necessary since three ferromagnets have different band structures. Fig.~\ref{fig:spin} shows that the inclusion of intraband transitions to the {QLE} (\ref{eq:QLE3}) dramatically enhances the demagnetization from 1-5\% in (a) to 50-90\% in (b).  This is in accordance with the experiments where moderate fluences ($\sim 10$ mJ/cm$^2$) were sufficient to demagnetize ferromagnets more than 50\% \cite{carpene2008,roth2012,you2018,borchert2021}.  Previous ab-initio studies using the velocity gauge, based either on the time-dependent density functional theory \cite{krieger2015,elliott2016b,siegrist2019,scheid2021,dewhurst2021b} or the time-dependent Liouville density functional theory \cite{jap09,jpcm16}, were unable to achieve demagnetization of ferromagnets greater than 10\% when using the experimental fluence, and that is most likely because they did not explicitly include the intraband transitions in their calculations. We also find that  a linearly-polarized pulse works as well, as shown with dash-dotted line in Fig.~\ref{fig:spin}(b).

In our bulk materials, without interfaces or surfaces, the spin moment is reduced due to the transition from the $3d$ bands with a larger spin moment to the $4sp$ bands with a smaller spin moment.  Because the $3d$ bands have a stronger exchange interaction, the spin majority and minority bands split, but in the excited states the $4dp$ bands are weakly spin-polarized.  The presence of spin-orbit coupling (SOC) mixes the minority and majority spin states, and allows the spin flipping transitions, which reduces the total spin moment.  In the absence of SOC, the spin moment cannot be reduced in this picture.  There are other mechanisms for demagnetization without SOC.  One is to allow the spin to transport from one part of the sample to another.  The part losing the majority electron sees the decrease in the spin moment.  The receiving end sees the increase in spin moment.    Another one is the spin wave excitation \cite{jap19}.  

\subsection{Bracket energy dependence}

\begin{figure}[!h]
\centering
\includegraphics[width=.95\textwidth, trim = 0in 0in 0in 0in, clip]{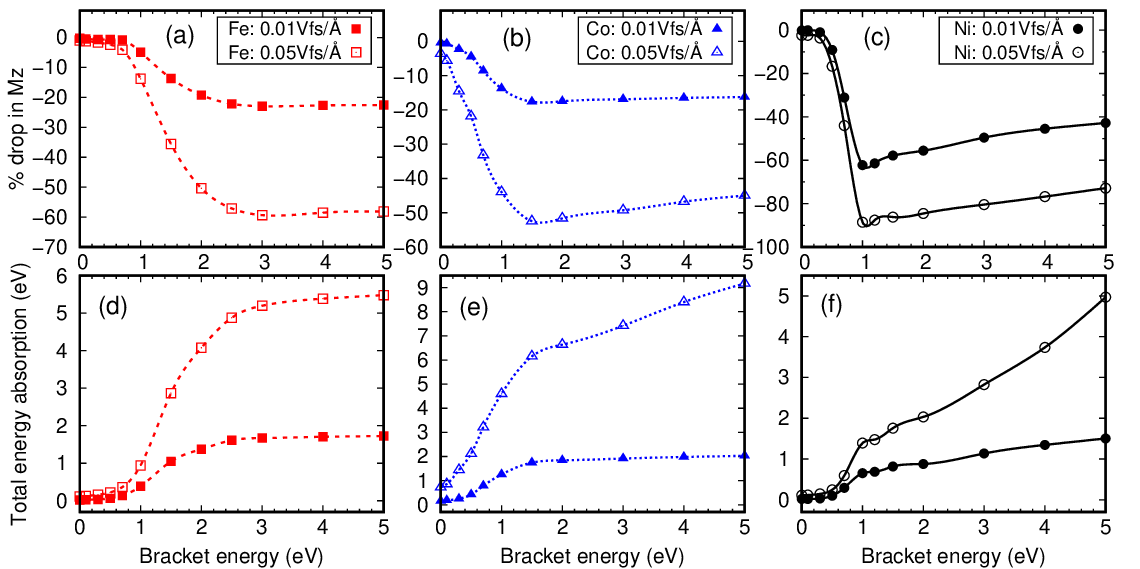}
\caption{(a)-(c): The percent reduction in the amount of magnetic moment at the end of  a circularly-polarized $\omega=1.6$ eV (775 nm) laser pulse of duration $\tau=60$ fs, as a function of bracket energies $E_b$ in Eq.~(\ref{eq:Ederiv}).  For each ferromagnetic sample (bcc Fe, hcp Co and fcc Ni), results with two different peak vector potentials are shown: $A_0=0.01$ V$\cdot$fs/{\AA}  and  0.05 V$\cdot$fs/{\AA}.  The corresponding fluences are $F_0= \sqrt{\frac{\pi}{2}}\frac{c\epsilon_0}{2}(\omega A_0)^2 = 0.587$ mJ/cm$^2$ and 14.7 mJ/cm$^2$, respectively.  (d)-(f): The total absorbed energy by electrons at the end of the laser pulse. }
\label{fig:Mz}
\end{figure}

Figures \ref{fig:Mz}(a)-(c) show the amount of demagnetization at the end of a laser pulse, $M_z(\infty)/M_0-1$, as a function of bracket energy $E_b$ in Eq.~(\ref{eq:Ederiv}).  The bracket energy is a parameter we use to control 
whether a state enters the intraband transition calculation.    In particular, zero bracket energy means no intraband transition.   For each sample (bcc Fe, hcp Co and fcc Ni), results with two different peak vector potentials are shown, $A_0=0.01$ V$\cdot$fs/{\AA} and $0.05$ V$\cdot$fs/{\AA}, whose corresponding incident fluences are $F_0= \sqrt{\frac{\pi}{2}}\frac{c\epsilon_0}{2}(\omega A_0)^2 = 0.587$ mJ/cm$^2$ and 14.7 mJ/cm$^2$, respectively, comparable to the fluence used in a recent experiment (0.5-15 mJ/cm$^2$) of Ref.~\cite{borchert2021}.  There is a common trend with both fluences in Figs \ref{fig:Mz}(a)-(c), namely, the magnetization decreases first as the amount of bracket energy increases, and then either stabilizes (for bcc Fe) or starts to increase slightly (for fcc Ni, and hcp Co when driving-laser vector potential is stronger).  The largest magnetization reduction of each material in Fig.~\ref{fig:Mz}(a) occurs when $E_b=$ 3 eV for Fe, 1.5 eV for Co and 1 eV for Ni.  Extension of bracket energy beyond these cutoff values does not contribute to further demagnetization.   

For bcc Fe, the amount of demagnetization in Fig.~\ref{fig:Mz}(a) stabilizes at -20\% and -60\% for $A_0=0.01$ V$\cdot$fs/{\AA} and $0.05$ V$\cdot$fs/{\AA}, respectively, once the bracket energy reaches its cutoff value.  On the other hand, the loss of magnetization in fcc Ni in Fig.~\ref{fig:Mz}(c) peaks at the cutoff (1 eV), beyond which the final magnetization slightly increases with the bracket energy for both fluences.  Similar increase is observed with hcp Co in Fig.~\ref{fig:Mz}(b) as well, but only for the higher fluence (open triangles).    Since we do not include the effect of dephasing due to electron-phonon scattering in our present calculation, it is not obvious what causes this increase in final magnetization when we allow the bracket energy to exceed the cutoff value.  One possible explanation is that the driving laser is causing the ringing of spin magnetic moments, which is known to make the spin magnetic moment to flip back to the initial state \cite{gerrits2002}.   To find out if the ringing is in fact causing the increase of final magnetization, we plot the total energy absorption by electrons as a function of bracket energy in Figs \ref{fig:Mz}(d)-(f).  Notice that the change in the final magnetization of hcp Co and fcc Ni beyond the cutoff energy in Figs \ref{fig:Mz}(b) and (c) is associated with the increase of absorbed energy in Figs \ref{fig:Mz}(e) and (f).  This supports the idea that absorption of extra energy beyond the cutoff (i.e., the maximum energy an electron could gain by intraband transitions) may cause the ringing and disturb demagnetization.   Fig.~\ref{fig:Mz} further suggests that higher fluence is more prone to cause the ringing in Ni and Co, whereas Fe is not susceptible to the ringing. This is reasonable because Fe has fewer valence electrons than Co or Ni.   Another striking observation in Fig.~\ref{fig:Mz} is that the hcp Co crystal is least efficient in converting the absorbed energy into demagnetization.  The cutoff energy of hcp Co (1.5 eV) is near resonant with the driving laser (1.6 eV), which explains the largest energy absorption of hcp Co in Fig.~\ref{fig:Mz}(e), but the resulting demagnetization of hcp Co in Fig.~\ref{fig:Mz}(b) is the smallest of all three samples, consistent with the experimental observation of pure Ni, Fe and Co films in Ref.~\cite{borchert2021}.  It is fascinating to find such diverse responses from different ferromagnets under the same laser pulses in Fig.~\ref{fig:Mz}, reflecting on their underlying band structures.  

\subsection{Band structure and cutoff bracket energy}

\begin{figure*}[!h]
	\centering
	\begin{minipage}{.31\textwidth}
		\includegraphics[width=\textwidth, trim = .02in 0in .02in 0in, clip]{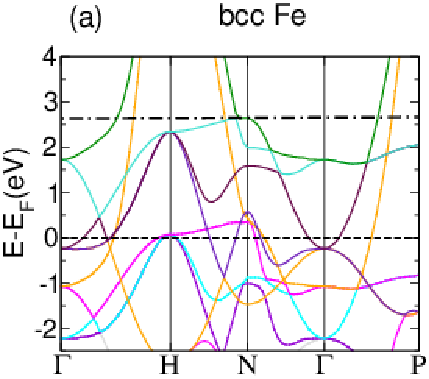}
	\end{minipage}
	\begin{minipage}{.31\textwidth}
		\includegraphics[width=\textwidth, trim = .02in 0in .03in 0in, clip]{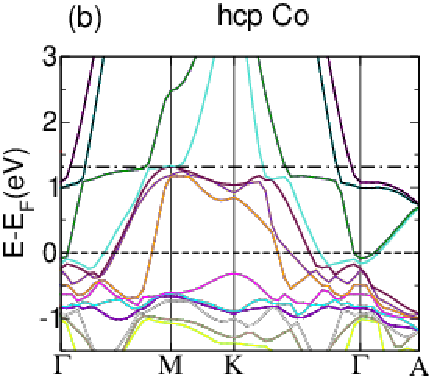}
	\end{minipage}
	\begin{minipage}{.31\textwidth}
		\includegraphics[width=\textwidth, trim = .02in 0in .04in 0in, clip]{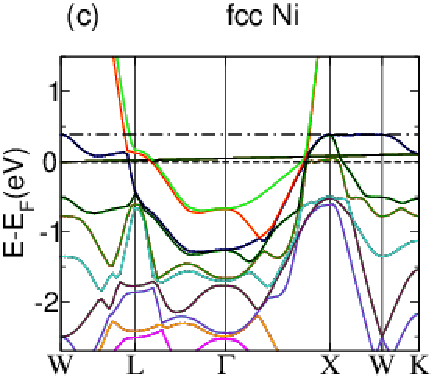}
	\end{minipage}
	\caption{Band structure of (a) {bcc Fe}, (b) hcp Co and (c) fcc Ni.  For each crystal, a dash-dotted line is drawn at the peak of the $3d$ band, and a dash line is the Fermi level at 0 eV.  {We find that} the depth of the $3d$ band at the $\Gamma$ point measured from the dash-dotted line is roughly 1 eV, 3 eV and 1.5 eV, respectively, corresponding to the cutoff bracket energies which give the largest magnetization reduction in Fig.~\ref{fig:Mz}(a).  }
\label{fig:band}
\end{figure*}

The fact that the cutoff bracket energies  in Fig.~\ref{fig:Mz} are independent of driving-laser fluence suggests that their origin is not dynamical but structural.  Therefore, we plot the band structures of bcc Fe, hcp Co and fcc Ni in Fig.~\ref{fig:band}.  Observation of Fig.~\ref{fig:band} reveals that cutoff bracket energy in Fig.~\ref{fig:Mz}(a)-(c) corresponds to the depth of a $3d$ band in each ferromagnetic crystal.  That is, maximum demagnetization in each ferromagnet is achieved when an electron gains enough energy to move up to the top of a $3d$ band via intraband transitions.  
Notice in Fig.~\ref{fig:band}(a) that the $3d$ band of bcc Fe crosses the Fermi level (horizontal dashed line) and ends around 2.6 eV (dash-dotted line).  Measured from the bottom of the $3d$ band at the $\Gamma$ point, this energy naturally sets the upper limit for our bracket energy.  The band structures for hcp Co and fcc Ni in Fig.~\ref{fig:band}(b) and (c) have shallower $3d$ bands than bcc Fe.  With one and two more electrons, the available number of $3d$ states is reduced.  This explains why the cutoff values in the bracket energy of Fig.~\ref{fig:Mz} are the largest for Fe (3 eV), 1.5 eV for Co, and the smallest for Ni (1 eV).

The above analysis indicates that we should be cautious about setting an arbitrary large bracket energy in Eq.~(\ref{eq:Ederiv}).  In fact, the bracket energy in our calculation ought to be determined by the depth of the highly localized $3d$ band of each metal.  Using a too large bracket energy involves the highly dispersive $4sp$ band and should be avoided because the intraband transition is not supposed to change the band character. After all, the intraband transition is electron dynamics within the band itself. The successful treatment  of intraband contribution to demagnetization relies on the proper treatment of $3d$ states. These are the same states that real space spin transport depends on.  

\begin{figure*}[!h]
\centering
\includegraphics[width=\textwidth, trim = 0in 0in 0in 0in, clip]{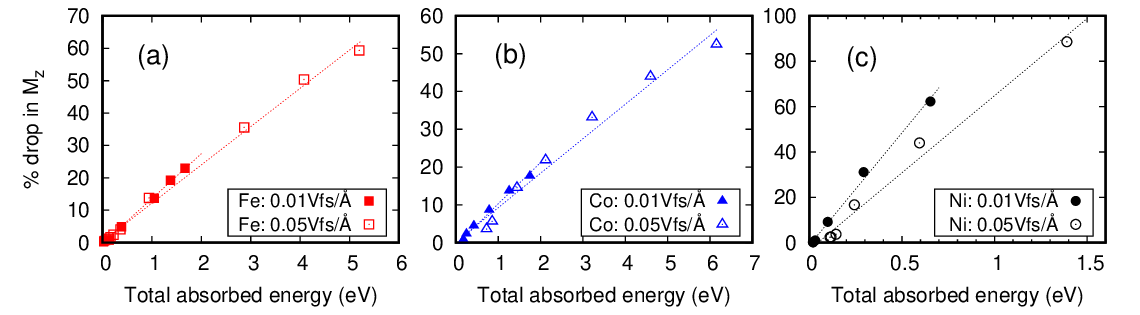}
\caption{(Color online)  Correlation plot between the total absorbed energy and the magnetization reduction of (a) bcc Fe, (b) hcp Co and (c) fcc Ni for various bracket energies in Fig.~\ref{fig:Mz}. The bracket energy for each ferromagnetic crystal is restricted below their cutoff value in order to avoid highly dispersive $4sp$ bands.}
\label{fig:corr}
\end{figure*}

In order to further study the effect of intraband transitions to demagnetization, we plot in Fig.~\ref{fig:corr} the total absorbed energy against the percent magnetization reduction of Fig.~\ref{fig:Mz} for various bracket energies. The bracket energy in Fig.~\ref{fig:corr} is restricted below the cutoff values to avoid highly dispersive $4sp$ bands.    There is a clear positive correlation between the two variables in Fig.~\ref{fig:corr}.  Moreover, we find that the amount of energy gained from the intraband transition does not universally determine the amount of demagnetization; rather, its efficiency for demagnetization depends on atomic species.  For example, the unoccupied $3d$ band of bcc Fe in Fig.~\ref{fig:band} is three times deeper than fcc Ni in Fig.~\ref{fig:band}, but bcc Fe in Fig.~\ref{fig:Mz}(a) does not demagnetize as much as fcc Ni in Fig.~\ref{fig:Mz}(c).   For bcc Fe and hcp Co, linear regression of Fig.~\ref{fig:corr} gives 13\% and 10\% magnetization reduction per eV of absorbed energy, respectively, whereas fcc Ni demagnetizes more than 80\%.

\subsection{Comparison with prior experiments and theories}

\begin{figure*}[!h]
	\centering
	\begin{minipage}{.45\textwidth}
		\includegraphics[width=\textwidth, trim = 0in 0in 0.05in 0in, clip]{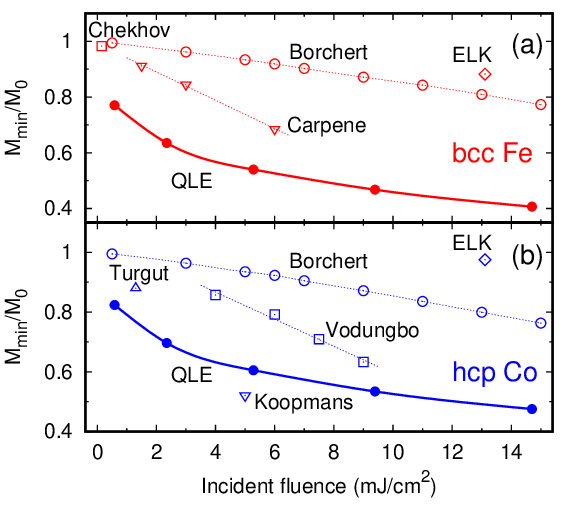}
	\end{minipage}
	\begin{minipage}{.45\textwidth}
		\includegraphics[width=\textwidth, trim = 0in 0in 0.05in 0in, clip]{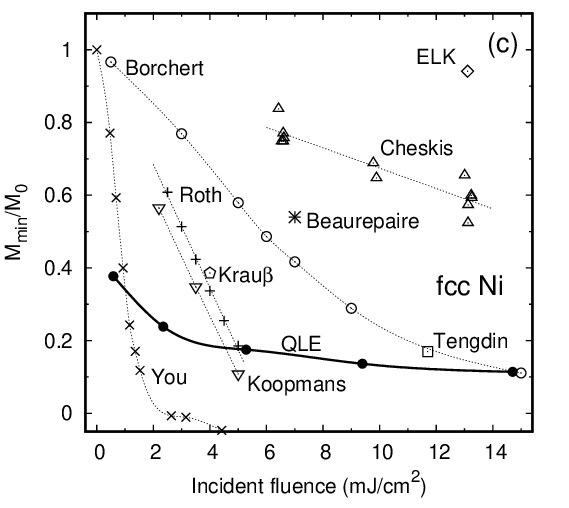}
	\end{minipage}
	\caption{Normalized magnetization at their minimum (before remagnetization starts due to electron-phonon scattering) of (a) bcc Fe, (b) hcp Co and (c) fcc Ni as a function of incident fluence.  Solid lines are our solutions of the quantum Liouville equation (QLE) with the cutoff bracket energy ($E_b=1$ eV, 3 eV and 1.5 eV, respectively), and dotted lines are experimental results of Refs.~\cite{cheskis2005,carpene2008,koopmans2010,roth2012,vodungbo2012,you2018,borchert2021}.  The 1.6 eV (775 nm) driving laser pulse in the QLE calculation has a duration of $\tau=60$ fs.  Note that the results of You et al.~in Ref.~\cite{you2018} are given in terms of {\it absorbed} fluence which should be several times smaller than incident fluence.  Also shown are the TDDFT calculation based on the ELK code in Ref.~\cite{scheid2021} and the experimental results of Refs.~\cite{eric,krauss2009,chechov2021,turgut2016,tengdin2018}.}
\label{fig:fluence}
\end{figure*}

Finally in Fig.~\ref{fig:fluence}, we plot the amount of maximum demagnetization in each ferromagnetic sample predicted by the QLE for various incident fluences.   Also shown for comparisons are experimental results \cite{eric,cheskis2005,carpene2008,krauss2009,koopmans2010,roth2012,vodungbo2012,turgut2016,tengdin2018,you2018,chechov2021,borchert2021} and the time-dependent density functional theory (TDDFT) calculations by the ELK software in Ref.~\cite{scheid2021}.  Direct comparisons with experiments are difficult because measurements of magnetization are influenced by the thickness of crystals \cite{you2018} as well as the temperature \cite{roth2012}. Under the assumption of linear absorption, the in-situ laser fluence decays exponentially as it penetrates through a crystal \cite{you2018}.  Therefore, demagnetization measured in experiments is expected to be less effective than our theoretical calculation predicts.  This is probably why our QLE calculations tend to demagnetize better than experiments do, particularly when the incident fluence is below 10 mJ/cm$^2$.  It is contrasting that the QLE predicts larger demagnetization than most experiments whereas the ELK calculations tend to predict less in Fig.~\ref{fig:fluence}.  This is expected, since we maximize the demagnetization due to intraband transitions by letting the bracket energy in Eq.~(\ref{eq:Ederiv}) be the cutoff value when solving the QLE for these figures.  The ELK calculation (which is based on the velocity gauge, similarly to ours), on the other hand,  do not explicitly incorporate the intraband transitions, which underestimates the amount of induced dipole moment and thus demagnetization.  Fig.~\ref{fig:fluence} shows that the effect of intraband transition cannot be ignored when calculating the demagnetization of ferromagnetic crystals.

\section{\label{sec:conc} Conclusion} 

We have shown that intraband transitions have a significant effect on
demagnetization in all three transition-element ferromagnets.  The
contribution of intraband transitions to demagnetization in each
element is different because of the band structure difference. The
amount of spin moment reduction, under the same experiment laser
fluence, is now comparable to the experimental data.  Our method is
based on the time-dependent Liouville equation using the velocity
gauge, which is advantageous over the length gauge where different
crystal momenta are coupled to each other.  In this work, we only
address the maximum demagnetization and in the future we plan to
investigate the time profile of demagnetization, so we can compare
with experiments to the full extent \cite{carpene2008}.  Our work
paves the way to fully understand laser-induced ultrafast
demagnetization.

\begin{acknowledgments}
M.~M.~thanks Drs.~J.~P.~Colgan and A.~B.~Saxena at Los Alamos National
Laboratory for their hospitality and a guest arrangement.  This work
was supported by the U.S. Department of Energy under Contract
No.~DE-FG02-06ER46304.  Numerical calculation was done on Indiana
State University's quantum cluster and high-performance computer
(obsidian).  The research used resources of the National Energy
Research Scientific Computing Center, which is supported by the Office
of Science of the U.S. Department of Energy under Contract
No.~DE-AC02-05CH11231.
\end{acknowledgments}

$^*$ guo-ping.zhang@outlook.com.
 https://orcid.org/0000-0002-1792-2701


\begin{thebibliography}{99}
\bibitem{ourbook}G. P. Zhang, G. Lefkidis, M. Murakami, W.  H\"ubner,
and T. F. George, {\it Introduction to Ultrafast Phenomena: From
Femtosecond Magnetism to High-Harmonic Generation}, CRC Press,
Taylor \& Francis Group, Boca Raton, Florida  (2021).

\bibitem{eric} E. Beaurepaire, J. C. Merle, A. Daunois, and J.-Y. Bigot,
{Ultrafast spin dynamics in ferromagnetic nickel},
Phys.  Rev. Lett. {\bf 76}, 4250 (1996).

\bibitem{vaterlaus1991} A. Vaterlaus, T. Beutler, and F. Meier,
{Spin-lattice relaxation time of ferromagnetic gadolinium determined
with time-resolved spin-polarized photoemission},
Phys. Rev. Lett. {\bf 67}, 3314 (1991).

\bibitem{koopmans2010}B. Koopmans, G. Malinowski, F. Dalla Longa,
D. Steiauf, M. F\"ahnle, T. Roth, M. Cinchetti, and M. Aeschlimann,
{Explaining the paradoxical diversity of ultrafast laser-induced
demagnetization}, Nat. Mater.  {\bf 9}, 259 (2010).

\bibitem{rasingreview}A. Kirilyuk, A. V. Kimel, and Th. Rasing,
{Ultrafast optical manipulation of magnetic order},
Rev. Mod. Phys. {\bf 82}, 2731 (2010). Erratum: Rev. Mod. Phys. {\bf
88}, 039904 (2016).

\bibitem{mplb16}G. P. Zhang, T. Latta, Z. Babyak, Y. H. Bai, and
T. F. George, {All-optical spin switching: A new frontier in
femtomagnetism --  A short review and a simple theory}, Mod. Phys.
Lett. B  {\bf 30}, 1630005 (2016).

\bibitem{prl20} S. R. Acharya, V. Turkowski, G. P. Zhang, and
T. Rahman, {Ultrafast electron correlations and memory effects at
work: Femtosecond demagnetization in Ni}, Phys. Rev. Lett. {\bf    125}, 017202 (2020).

\bibitem{kimel2019}A. V. Kimel and M. Li, {Writing magnetic memory
with ultrashort light pulses}, Nat. Rev. Mater. {\bf 4}, 189
(2019).

\bibitem{scholl1997}A. Scholl, L. Baumgarten, R. Jacquemin, and
W. Eberhardt, {Ultrafast spin dynamics of ferromagnetic thin films
observed by fs spin-resolved two-photon photoemission},
Phys. Rev. Lett. {\bf 79}, 5146 (1997).

\bibitem{prl00} G. P. Zhang and W. H\"ubner, {Laser-induced ultrafast
demagnetization in ferromagnetic metals}, Phys. Rev. Lett. {\bf 85},
3025 (2000).

\bibitem{gerrits2002} Th. Gerrits, H. A. M. van den Berg, J. Hohlfeld,
L. B\"ar and Th. Rasing, {Ultrafast precessional magnetization
reversal by picosecond magnetic field pulse shaping}, Nature {\bf
418}, 509 (2002).

\bibitem{ramsay}A. J. Ramsay, P. E. Roy, J. A. Haigh, R. M. Otxoa,
A. C. Irvine, T. Janda, R. P. Campion, B. L. Gallagher, and
J. Wunderlich, {Optical spin-transfer-torque-driven domain-wall
motion in a ferromagnetic semiconductor}, Phys. Rev. Lett. {\bf  114}, 067202 (2015).

\bibitem{lan2017} J. Lan, W. Yu and J. Xiao, {Antiferromagnetic domain
wall as spin wave polarizer and retarder}, Nat. Commun. {\bf 8}, 178
(2017).

\bibitem{rhie2003}H.-S. Rhie, H. A. D\"urr, and W. Eberhardt, {Femtosecond
electron and spin dynamics in Ni/W(110) films}, Phys. Rev. Lett. {\bf   90}, 247201 (2003).

\bibitem{muller2009}G. M. M\"uller, J. Walowski, M.  Djordjevic,
G.-X. Miao, A. Gupta, A. V. Ramos, K. Gehrke, V. Moshnyaga,
K. Samwer, J. Schmalhorst, A. Thomas, A. H\"utten, G.  Reiss,
J. S. Moodera, and M. M\"unzenberg, {Spin polarization in
half-metals probed by femtosecond spin excitation},
Nat. Mater. {\bf 8}, 56 (2009).

\bibitem{kimel2005}A. V. Kimel, A. Kirilyuk, P. A. Usachev,
R. V. Pisarev, A. M. Balbashov, and Th. Rasing, {Ultrafast
non-thermal control of magnetization by instantaneous
photomagnetic pulses}, Nature {\bf 435}, 655 (2005).

\bibitem{steiauf2009}D. Steiauf and M. Fähnle, {Elliott-Yafet
mechanism and the discussion of femtosecond magnetization dynamics},
Phys. Rev. B {\bf 79}, 140401(R) (2009).

\bibitem{krauss2009}M. Krau\ss, T. Roth, S. Alebrand, D. Steil, M.
Cinchetti, M. Aeschlimann, and H. C. Schneider, {Ultrafast
demagnetization of ferromagnetic transition metals: The role of
the Coulomb interaction}, Phys.  Rev. B {\bf 80}, 180407(R)
(2009).

\bibitem{schmidt2010}A. B. Schmidt, M. Pickel, M. Donath, P. Buczek,
A. Ernst, V. P. Zhukov, P. M. Echenique, L. M. Sandratskii,
E. V. Chulkov, and M. Weinelt, {Ultrafast magnon generation in an Fe
film on Cu(100)}, Phys. Rev. Lett. {\bf 105}, 197401 (2010).

\bibitem{haag2014}M. Haag, C. Illg, and M. F\"ahnle, {Role of
electron-magnon scatterings in ultrafast demagnetization},
Phys. Rev. B {\bf 90}, 014417 (2014).

\bibitem{battiato2010} M. Battiato, K. Carva, and P. M. Oppeneer,
{Superdiffusive spin transport as a mechanism of ultrafast
demagnetization}, Phys. Rev. Lett. {\bf 105}, 027203 (2010).

\bibitem{schellekens2013}A. J. Schellekens, W. Verhoeven, T. N. Vader,
and B. Koopmans, {Investigating the contribution of superdiffusive
transport to ultrafast demagnetization of ferromagnetic thin
films},  Appl. Phys. Lett. {\bf 102}, 252408 (2013).

\bibitem{ostler2012}T. A. Ostler, J. Barker, R. F. L. Evans,
R. W. Chantrell, U. Atxitia, O. Chubykalo-Fesenko, S. El Moussaoui,
L. Le Guyader, E. Mengotti, L. J. Heyderman, F. Nolting,
A. Tsukamoto, A. Itoh, D. Afanasiev, B. A. Ivanov,
A. M. Kalashnikova, K. Vahaplar, J. Mentink, A. Kirilyuk,
Th. Rasing, and A. V. Kimel, {Ultrafast heating as a sufficient
stimulus for magnetization reversal in a ferrimagnet},
Nat. Commun. {\bf 3}, 666 (2012).

\bibitem{dornes2019}C. Dornes {\it et al.}, {The ultrafast
Einstein-de Haas effect}, Nature {\bf 565}, 209 (2019).

\bibitem{turgut2016} E. Turgut, D. Zusin, D. Legut, K. Carva, R. Knut,
J. M. Shaw, C. Chen, Z. Tao, H. T. Nembach, T. J. Silva, S. Mathias,
M. Aeschlimann, P. M. Oppeneer, H. C. Kapteyn, M. M. Murnane and
P. Grychtol, {Stoner versus Heisenberg: Ultrafast exchange reduction
and magnon generation during laser-induced demagnetization},
Phys. Rev. B {\bf 94}, 220408 (2016).

\bibitem{choi2017} G.-M. Choi, A. Schleife, and D. G. Cahill,
{Optical-helicity-driven magnetization dynamics in metallic
ferromagnets},  Nat. Commun. {\bf 8}, 15085  (2017).

\bibitem{buhlmann2018} K. B\"uhlmann, R. Gort, G. Salvatella,
S. Däster, A. Fognini, T. Bähler, C. Dornes, C. A. F. Vaz,
A. Vaterlaus and Y. Acremann, {Ultrafast demagnetization in iron:
Separating effects by their nonlinearity}, Struct. Dyn. {\bf 5},
044502 (2018).

\bibitem{scheid2021} P. Scheid, S. Sharma, G.  Malinowski, S. Mangin
and S. Lebegue, {Ab Initio Study of Helicity-Dependent Light-Induced
Demagnetization: From the Optical Regime to the Extreme
Ultraviolet Regime}, Nano Lett. {\bf 21}, 1943 (2021).

\bibitem{krieger2015} K. Krieger, J. K. Dewhurst, P. Elliott,
S. Sharma, and E. K. U. Gross, {Laser-induced demagnetization at
ultrashort time scales: Predictions of TDDFT}, J. Chem. Theory and
Comput. {\bf 11}, 4870 (2015).

\bibitem{elliott2016b} P Elliott, K. Krieger, J. K. Dewhurst,
S. Sharma and E. K. U. Gross, {Optimal control of laser-induced
spin–orbit mediated ultrafast demagnetization},
New J. Phys. {\bf 18},  013014 (2016).

\bibitem{siegrist2019}F. Siegrist, J. A. Gessner, M.  Ossiander,
C. Denker, Y.-P. Chang, M. C. Schr\"oder, A. Guggenmos, Y. Cui,
J. Walowski, U. Martens, J. K. Dewhurst, U. Kleineberg,
M. M\"unzenberg, S.  Sharma, and M. Schultze, {Light-wave dynamic
control of magnetism}, Nature {\bf 571}, 240 (2019).

\bibitem{dewhurst2021b}J. K. Dewhurst, S. Shallcross, P. Elliott,
S. Eisebitt, C. v. Korff Schmising, and S. Sharma, {Angular momentum
redistribution in laser-induced demagnetization}, Phys. Rev. B
{\bf 104}, 054438 (2021).

\bibitem{jpcm16}G. P. Zhang, Y. H. Bai, and T. F. George, {Ultrafast
reduction of exchange splitting in ferromagnetic nickel}, J. Phys.:
Condens. Matter {\bf 28}, 236004 (2016).

\bibitem{you2018}W. You, P. Tengdin, C. Chen, X. Shi, D. Zusin,
Y. Zhang, C. Gentry, A. Blonsky, M. Keller, P. M. Oppeneer,
H. Kapteyn, Z. Tao, and M. Murnane, {Revealing the nature of the
ultrafast magnetic phase transition in Ni by correlating extreme
ultraviolet magneto-optic and photoemission spectroscopies},
Phys. Rev. Lett. {\bf 121}, 077204 (2018).

\bibitem{aversa1995}C. Aversa and J. E. Sipe, {Nonlinear optical
susceptibilities of semiconductors: Results with a length-gauge
analysis}, Phys. Rev. B   {\bf 52}, 14636 (1995).

\bibitem{golde2008} D. Golde, T. Meier and S. W. Koch, {High harmonics
generated in semiconductor nanostructures by the coupled dynamics of
optical inter- and intraband excitations}, Phys. Rev. B {\bf 77},
075330 (2008).

\bibitem{blount1962}E. I. Blount, {Formalisms of Band Theory}, Solid
State Phys. {\bf 13}, 305 (1962).

\bibitem{krieger1986} J. B. Krieger and G. J. Iafrate, {Time
evolution of Bloch electrons in a homogeneous electric field},
Phys. Rev. B {\bf 33}, 5494 (1986).

\bibitem{wu2015} M. Wu, S. Ghimire, D. A. Reis, K. J. Schafer, and
M. B. Gaarde, {High harmonic generation from Bloch electrons in
solids}, Phys. Rev. A {\bf 91}, 043839 (2015).

\bibitem{np09}G. P. Zhang, W. H\"ubner, G. Lefkidis, Y. Bai, and
T. F. George, {Paradigm of the time-resolved magneto-optical Kerr
effect for femtosecond magnetism}, {Nat. Phys.} {\bf 5}, 499
(2009).

\bibitem{koopmans2000}B. Koopmans, M. van Kampen, J. T. Kohlhepp, and
W. J. M. de Jonge, {Ultrafast magneto-optics in nickel: magnetism or
optics?}  Phys. Rev. Lett. {\bf 85}, 844 (2000).

\bibitem{ghimire2011} S. Ghimire, E. Sistrunk, P. Agostini,
L. F. DiMauro, and D. A. Reis, {Observation of high-order harmonic
generation in a bulk crystal}, Nat. Phys. {\bf 7}, 138
(2011).

\bibitem{schubert2014} O. Schubert, M. Hohenleutner, F. Langer,
B. Urbanek, C. Lange, U. Huttner, D. Golde, T. Meier, M. Kira,
S. W. Koch and R. Huber, {Sub-cycle control of terahertz
high-harmonic generation by dynamical Bloch oscillations},
Nat. Photon. {\bf 8}, 119 (2014).

\bibitem{luu2015}T. T. Luu, M. Garg, S. Yu. Kruchinin, A. Moulet,
M. Th. Hassan, and E. Goulielmakis, {Extreme ultraviolet
high-harmonic spectroscopy of solids}, Nature {\bf 521}, 498 (2015).

\bibitem{vampa2014}G. Vampa, C. R. McDonald, G. Orlando, D. D. Klug,
P. B. Corkum and T. Brabec, {Theoretical analysis of high-harmonic
generation in solids}, Phys. Rev. Lett. {\bf 113}, 073901 (2014).

\bibitem{vampa2015} G. Vampa, T. J. Hammond, N. Thiré, B. E. Schmidt,
F. Légaré, C. R. McDonald, T. Brabec and  P. B. Corkum, {Linking high
harmonics from gases and solids}, Nature {\bf 522},  462 (2015).

\bibitem{kittel2}C. Kittel, {\it Quantum theory of solids},
John Wiley \& Sons, Inc., New York (1964). Page 192.

\bibitem{mahan2000}G. D. Mahan, {\it Many-Particle Physics}, Springer,
Boston, MA (2000).

\bibitem{griffiths2013} D. J. Griffiths, {\it Introduction to
electrodynamics}, Pearson, page 443 (2013).

\bibitem{koelling1977}D. D. Koelling and B. N. Harmon, {A technique
for relativistic spin-polarised calculations}, J. Phys. C {\bf 10},
310 (1977).

\bibitem{perdew2003}J.P. Perdew and S. Kurth, {Density Functionals for
Non-relativistic Coulomb Systems in the New Century} in {Lecture
Notes in Physics}, {C. Fiolhais, F. Nogueira, and M.A.L.  Marques}
ed. Springer, Berlin, Heidelberg (2003).

\bibitem{blaha2010}P. Blaha, H. Hofst\"atter, O. Koch, R. Laskowski
and K. Schwarz, {Iterative Diagonalization in Augmented Plane Wave
Based Methods in Electronic Structure Calculations},
J. Comput. Phys. {\bf 229}, 453 (2010).

\bibitem{blaha2020}P. Blaha, K. Schwarz, F. Tran, R. Laskowski,
G. K. H. Madsen and L. D. Marks, {WIEN2k: An APW+lo program for
calculating the properties of solids}, J. Chem. Phys. {\bf 152},
074101 (2020).

\bibitem{pbe}J. P. Perdew, K. Burke and M. Ernzerhof, {Generalized
gradient approximation made simple}, Phys. Rev. Lett. {\bf 77}, 3865
(1996).

\bibitem{macdonald1980}A. H. MacDonald, W. E. Picket and
D. D. Koelling, {A linearised relativistic augmented-plane-wave
method utilising approximate pure spin basis functions},
J. Phys. C {\bf 13}, 2675 (1980).

\bibitem{madsen2002}L. B. Madsen, {Gauge invariance in the interaction
between atoms and few-cycle laser pulses}, Phys. Rev. A {\bf 65},
053417 (2002).

\bibitem{baurer2005}D. Bauer, D. B. Milosevic, and W. Becker,
{Strong-field approximation for intense-laser–atom processes: The
choice of gauge}, Phys. Rev. A {\bf 72}, 023415 {2005}.

\bibitem{prb09} G. P. Zhang, Y. H. Bai, and T. F. George, {Energy- and
crystal momentum-resolved study of laser-induced femtosecond
magnetism}, Phys. Rev. B {\bf 80}, 214415 (2009).

\bibitem{draxl2006}C. Ambrosch-Draxl and O. Sofo, {Linear optical
properties of solids within the full-potential linearized augmented
planewave method}, Comp. Phys. Comm. {\bf 175}, 1 (2006).

\bibitem{korbman2013}M. Korbman, S. Yu. Kruchinin and V. S. Yakovlev,
{Quantum beats in the polarization response of a dielectric to
intense few-cycle laser pulses}, New J. Phys,{\bf 15},
013006 (2013).

\bibitem{cheng2019}J. L. Cheng, J. E. Sipe, S. W. Wu and Chunlei Guo,
{Intraband divergences in third order optical response of 2D
systems}, APL Photonics {\bf 4}, 034201 (2019).

\bibitem{luu2016}T. T. Luu and J. J. W\"orner, {High-order harmonic
generation in solids: A unifying approach}, Phys. Rev. B {\bf 94},
115164 (2016).

\bibitem{al-naib2014}I. Al-Naib, J. E. Sipe, and M. M. Dignam, {High
harmonic generation in undoped graphene: Interplay of inter- and
intraband dynamics}, Phys. Rev. B {\bf 90}, 245423 (2014).

\bibitem{kittel}C. Kittel, {\it Introduction to Solid State Physics},
7th Ed., John Wiley \& Sons, Inc., New York (1996).

\bibitem{ashcroft}N. W. Ashcroft and N. D. Mermin, {\it Solid State
Physics}, Cengage Learning; 1st Ed. (January 2, 1976).

\bibitem{peierls}R. Peierls, {\it Quantum Theory of Solids}, {Oxford
University Press}, (2001).

\bibitem{roth2012}T. Roth, A. J. Schellekens, S. Alebrand, O. Schmitt,
D. Steil, B. Koopmans, M. Cinchetti, and M. Aeschlimann,
{Temperature Dependence of Laser-Induced Demagnetization in Ni: A
Key for Identifying the Underlying Mechanism}, Phys. Rev. X {\bf
2}, 021006 (2012).

\bibitem{carpene2008}E. Carpene, E. Mancini, C. Dallera, M. Brenna,
E. Puppin, and S. De Silvestri, {Dynamics of electron-magnon
interaction and ultrafast demagnetization in thin iron films},
Phys. Rev. B {\bf 78}, 174422 (2008).

\bibitem{borchert2021}M. Borchert, C. von Korff Schmising, D. Schick,
D. Engel, S. Sharma, S. Shallcross and S. Eisebitt1, {Uncovering
the role of the density of states in controlling ultrafast spin
dynamics}, arXiv:2008.12612v2 (2021).

\bibitem{jap09}   T. Hartenstein, G. Lefkidis,
W. H\"ubner, G. P. Zhang, and Y. Bai, {Time-resolved and
energy-dispersed spin excitation in ferromagnets and clusters under
influence of femtosecond laser pulses}, J.  Appl. Phys.
{\bf 105}, 07D305 (2009).

\bibitem{jap19}G. P. Zhang, M. Murakami, Y. H. Bai, T. F. George, and
X. S. Wu, {Spin-orbit torque-mediated spin-wave excitation as an
alternative paradigm for femtomagnetism}, J. Appl. Phys. {\bf 126},
103906 (2019).

\bibitem{cheskis2005}D. Cheskis, A. Porat, L. Szapiro, O. Potashnik
and S. Bar-Ad, {Saturation of the ultrafast laser-induced
demagnetization in nickel}, Phys. Rev. B {\bf 72}, 014437 (2005).

\bibitem{vodungbo2012}B. Vodungbo, J. Gautier, G.  Lambert,
A. B. Sardinha, M. Lozano, S.  Sebban, M. Ducousso, W. Boutu, K. Li,
B.  Tudu, M. Tortarolo, R. Hawaldar, R. Delaunay, V. Lopez-Flores,
J. Arabski, C. Boeglin, H.  Merdji, P. Zeitoun, and J. L\"uning,
{Laser-induced ultrafast demagnetization in the presence of a
nanoscale magnetic domain network}, Nat. Commun. {\bf 3}, 999
(2012).

\bibitem{tengdin2018}P. Tengdin, W. You, C. Chen, X.  Shi, D. Zusin,
Y. Zhang, C. Gentry, A.  Blonsky, M. Keller, P. M. Oppeneer,
H. C. Kapteyn, Z. Tao, and M. M. Murnane, {Critical behavior within
20 fs drives the out-of-equilibrium laser-induced magnetic phase
transition in nickel}, Sci. Adv. {\bf 4}, eaap9744 (2018).

\bibitem{chechov2021}A. L. Chekhov, Y. Behovits, J. J. F. Heitz,
C. Denker, D. A. Reiss, M. Wolf, M. Weinelt, P. W. Brouwer,
M. M\"unzenberg and T. Kampfrath, {Ultrafast Demagnetization of Iron
Induced by Optical versus Terahertz Pulses}, Phys. Rev. X {\bf
11},  041055 (2021).

\end{thebibliography}

\newpage

\end{document}